\documentclass[10pt,a4paper]{article}

\usepackage{times}
\usepackage{setspace} 

\usepackage{epsfig,wrapfig}
\usepackage{epstopdf} % for pdflatex

\usepackage{fancyhdr}

\usepackage[lmargin=2.5cm, rmargin=2.5cm,tmargin=3.50cm,bmargin=3.50cm]{geometry}
\usepackage{indentfirst}

\usepackage{titlesec}
\titleformat{\section}[hang]
  {\centering}{\thesection}{1ex}{\normalsize \textsc}%%
\titleformat{\subsection}[hang]
  {}{\thesubsection}{1ex}{\normalsize \textit}%%

\renewcommand{\thesection}{ \normalsize \textnormal{\Roman{section}.}}
\renewcommand{\thesubsection}{\normalsize \textnormal{\textsc{\textit{\Alph{subsection}.}}}}

\def\e{\begin{equation}}
\def\f{\end{equation}}
\def\_#1{{\bf #1}}

\def\.{\cdot}

\usepackage{amsmath}
\usepackage[hidelinks]{hyperref}
\usepackage{xcolor}
\usepackage{physics}
\usepackage{float}
\usepackage{caption}
\usepackage{subcaption}

\newcounter{paticounter}

\begin{document}

%%% Title of paper
\title{\large \textbf{Electromagnetic Scattering at an Arbitrarily Accelerated Interface}}
%
%%% Author(s) and affiliation
\def\affil#1{\begin{itemize} \item[] #1 \end{itemize}}
\author{\normalsize \bfseries K. De Kinder and C. Caloz
}
\date{}
\maketitle
\thispagestyle{fancy} % header also to the first page
\vspace{-6ex}
\affil{\begin{center}\normalsize KU Leuven, Department of Electrical Engineering, Kasteelpark Arenberg 10, 3001, Leuven, Belgium \\
\href{mailto:klaas.dekinder@kuleuven.be}{klaas.dekinder@kuleuven.be}
\end{center}}

%%% Abstract
\begin{abstract}
\noindent \normalsize
\textbf{\textit{Abstract} \ \ -- \ \ 
%%% Start here with text of abstract
We present a general analytical solution to the problem of electromagnetic scattering at a one-dimensional \emph{arbitrarily accelerated} space-time engineered-modulation (ASTEM) interface in the subluminal regime. We show that such an interface fundamentally produces chirping, whose profile can be designed according to specifications. This work represents an important step in the development of ASTEM crystals and holds significant potential for applications in microwave and optical devices reliant on chirp-based functionalities.
}
\end{abstract}

\section{Introduction} 
    Space-Time Engineered-Modulation (STEM) metamaterials, referred to as STEMs, are dynamic structures that are formed by modulating some parameter (e.g., the refractive index) of a host medium \cite{Caloz2019a,Caloz2019b}. Notably, despite their dynamic nature, they involve no net transfer of matter. Moreover, they share many electrodynamic properties of moving media, including the Doppler shift~\cite{Caloz2022,Doppler1842} and the Fresnel-Fizeau drag~\cite{Fresnel1818,Huidobro2019}, while offering the advantage of comprising no moving parts. Finally, compared to their moving-matter counterparts, they provide easy access to relativistic velocities and feature more diverse physics. 
    \par
    In recent years, STEM research has predominantly focused on space-time metamaterials with uniform modulation velocity (e.g., \cite{Deck-Leger2022}), or \emph{uniform} STEMs (USTEMs)~\cite{Caloz2022}. However, extending the modulation to nonuniform patterns, and hence transforming USTEMs into \emph{accelerated} STEMs (ASTEMs), introduces greater diversity and opens avenues for novel systems and applications~\cite{Caloz2022}. Recently, related research efforts have been dedicated to the electrodynamics of ASTEMs with uniform acceleration~\cite{bahrami2022,Bahrami2023}.
    \par
    In this work, we remove the restriction of uniform acceleration and consider arbitrary acceleration. Specifically, we provide a general \emph{analytical} solution for the related scattering coefficients and develop a synthesis method to determine the trajectory of the interface corresponding to arbitrary chirping specifications. 

\section{Arbitrarily Accelerated Interface}\label{sec: Arbitrary Accelerated Interface}
\begin{figure}[h!]
    \centering
    \includegraphics[width=0.4\textwidth]{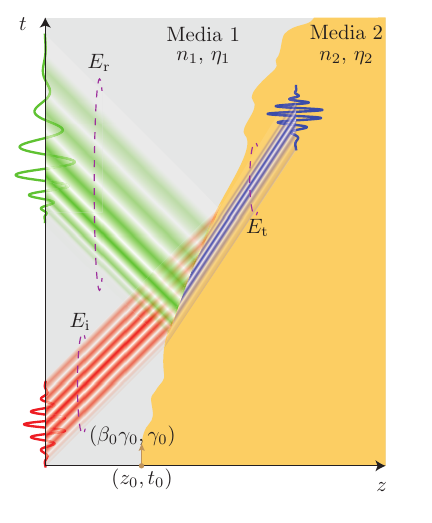}
    \caption{Spacetime scattering at an arbitrarily accelerated interface. The subscripts i, r and t stand for incident, reflected and transmitted, while the subscripts $1$ and $2$ refer to the two media.} 
    \label{fig: Example interface trajectory}
    \end{figure}

    We address the problem by leveraging the tools of special and general relativity, which enables us to conveniently represent the problem in the spacetime diagram shown in Fig.~\ref{fig: Example interface trajectory}. The interface is an arbitrary one-dimensional trajectory, $\left(z\left(\tau\right),t\left(\tau\right)\right)$ where $\tau$ is the proper time\footnote{The ``proper'' time is the time experienced by an observer moving along the trajectory of the interface.}, between two isotropic and nondispersive media of refractive indices $n_{1}$ and $n_{2}$ and impedances $\eta_{1}$ and $\eta_{2}$. It is characterized by (i) an initial space-time event, $(z_0,t_0)$, (ii) a normalized initial four-velocity, $\left(\gamma_{0}\beta_{0}, \gamma_{0}\right)$ where $\beta_{0}$ is the initial velocity of the interface and $\gamma_{0} = 1/\sqrt{1-\beta_{0}^{2}}$ the initial Lorentz factor, and (iii) a proper acceleration, $a\left(\tau\right)$ (curvature of the trajectory). An incident wave, $E_{\text{i}}$, impinges on the interface from the first medium. Upon encountering the interface, it undergoes scattering, resulting in a reflected wave, $E_{\text{r}}$, also propagating in the first medium and a transmitted wave, $E_{\text{t}}$, propagating in the second medium.   

\section{Electromagnetic Scattering Solutions}

    \subsection{Analysis Problem}
    The incident wave may be written as
    \begin{equation}\label{eq: Incident wave}
        E_{\text{i}} = A_{\text{i}} f\left(k_{\text{i}}z - \omega_{\text{i}}t\right) \,,
    \end{equation}
    where $f\left(\cdot\right)$ represents an arbitrary waveform profile. We subsequently apply the frame-hopping strategy~\cite{Bahrami2023,bladel2012}: the electromagnetic fields, $\boldsymbol{E}$ and $\boldsymbol{H}$, are transformed into their comoving-frame counterparts, where we apply the stationary boundary conditions, and the resulting complete fields are transformed back to the laboratory frame. After some algebraic manipulations, we obtain the following result\footnote{We adopt here natural units, where the speed of light is unity.}:
    \begin{subequations}\label{eq: Analytical results scattering coefficients}
        \begin{align}
            E_{\text{r}} = \frac{\eta_{2} - \eta_{1}}{\eta_{2} + \eta_{1}} \frac{1-n_{1}\beta\left(t\right)}{1+n_{1}\beta\left(t\right)} A_{\text{i}} \, f\left(\frac{1-n_{1}\beta\left(t\right)}{1+n_{1}\beta\left(t\right)}\left(k_{\text{i}}z + \omega_{\text{i}}t\right)\right) \,, \\
            E_{\text{t}} = \frac{2\eta_{2}}{\eta_{2}+\eta_{1}}\frac{1-n_{1}\beta\left(t\right)}{1-n_{2}\beta\left(t\right)} A_{\text{i}} \, f\left(\frac{1-n_{1}\beta\left(t\right)}{1-n_{2}\beta\left(t\right)}\left(k_{\text{i}}z - \omega_{\text{i}}t\right)\right)\,, \label{eq: Transmitted wave}
        \end{align}
        where $\beta\left(t\right)$ is the velocity of the interface as a function of the laboratory time and is given by
        \begin{equation}
            \beta\left(t\right) = \frac{\beta_{0}\gamma_{0} + \int_{0}^{t}a\left(t_{1}\right)\, \dd{t_{1}}}{\sqrt{1 + \left(\beta_{0}\gamma_{0} + \int_{0}^{t}a\left(t_{1}\right)\, \dd{t_{1}}\right)^{2}}} \,.
        \end{equation}
    \end{subequations}
    Surprisingly, the derivation of Eqs.~\eqref{eq: Analytical results scattering coefficients} does not require the explicit form of the coordinate transformation to the proper comoving frame: it turns out that the scattering coefficients in the moving frame can be expressed in terms of ratios of partial derivatives of the coordinate transformation, which ultimately obliterate the coordinate transformation.

    \subsection{Synthesis Problem}
\begin{figure}[h!]
        \centering
        \begin{subfigure}[b]{0.45\textwidth}
            \includegraphics[width=1\linewidth]{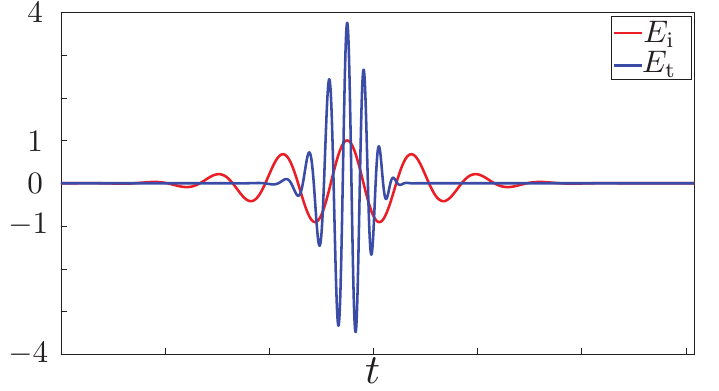}
            \caption{}
            \label{subfig: Scattered wave}
        \end{subfigure}
        \hfill
        \begin{subfigure}[b]{0.45\textwidth}
            \includegraphics[width=1\linewidth]{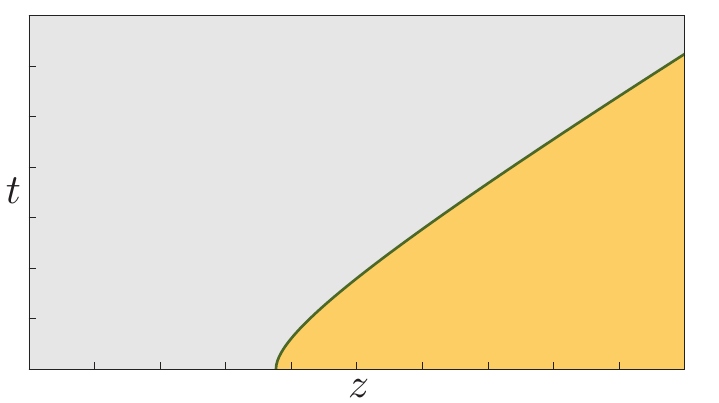}
            \caption{}
            \label{subfig: Spacetime diagram}
        \end{subfigure}
    \caption{Scattering at an accelerated modulated single interface with velocity in Eq.~\eqref{eq: Velocity for chirping} for the chirping profile in Eq.~\eqref{eq: Linear chirp} with chirp parameter $a = 5$ and the electromagnetic parameters $n_{1} = 1$, $n_{2} = 3$ and $\eta_{1} = 1 = \eta_{2}$. (a) Incident Gaussian modulated pulse [Eq.~\eqref{eq: Incident wave}] and transmitted wave [Eq.~\eqref{eq: Transmitted wave}]. (b) Trajectory of the interface [Eq.~\eqref{eq: Trajectory interface}] in a spacetime diagram.} 
    \label{fig: Validation}
    \end{figure}

    The fact that the phase argument of the expressions in Eqs.~\eqref{eq: Analytical results scattering coefficients} is a nonlinear function of time reveals that the scattered waves exhibit chirping when subjected to proper acceleration, as depicted in Fig.~\ref{fig: Example interface trajectory}. This prompts an inverse problem: determining the velocity of the interface for achieving a specified chirping profile in the scattered waves. To unravel this, say for the transmitted wave, we must solve the instantaneous angular frequency equation~\cite{Saleh2019},
    \begin{equation}
        \frac{1-n_{1}\beta\left(t\right)}{1-n_{2}\beta\left(t\right)}\omega_{\text{i}} = \omega_{\text{i}} + \varphi'\left(t\right)\,,
        \end{equation}
    with $\varphi\left(t\right)$ being the phase of the complex envelope of the wave, for the velocity $\beta\left(t\right)$, yielding the solution 
    \begin{equation}\label{eq: Velocity for chirping}
        \beta\left(t\right) = \frac{\varphi'\left(t\right)}{n_{2}\varphi'\left(t\right) + \left(n_{2}-n_{1}\right)\omega_{\text{i}}}\,.
    \end{equation}
    If we consider a linear upward chirp
    \begin{equation}\label{eq: Linear chirp}
        \varphi\left(t\right) = a \frac{t^{2}}{\alpha^{2}} \,,
    \end{equation}
    where $a > 0$ is the chirp parameter and $\alpha$ the width of the incident wave, the velocity profile [Eq.~\eqref{eq: Velocity for chirping}] can be integrated to
    \begin{equation}\label{eq: Trajectory interface}
        z\left(t\right) = z_{0} + \frac{t}{n_{2}} + \frac{\left(n_{2}-n_{1}\right)\omega_{\text{i}}\alpha^{2}}{2an_{2}^{2}} \log\frac{\left(n_{2}-n_{1}\right)\omega_{\text{i}}\alpha^{2}}{\left(n_{2}-n_{1}\right)\omega_{\text{i}}\alpha^{2} + 2an_{2}t}\,.
    \end{equation}
Figure~\ref{fig: Validation} illustrates the synthesis result for the linear upward chirp [Eq.~\eqref{eq: Linear chirp}] for $a = 5$, where the incident waveform [Eq.~\eqref{eq: Incident wave}] is a Gaussian modulated pulse (Fig.~\ref{subfig: Scattered wave}). The velocity of the interface is then obtained by inserting Eq.~\eqref{eq: Linear chirp} into Eq.~\eqref{eq: Velocity for chirping}, while the chirped transmitted wave is obtained by inserting the so-obtained velocity into Eq.~\eqref{eq: Transmitted wave}. Figure \ref{subfig: Spacetime diagram} plots the trajectory of the interface [Eq.~\eqref{eq: Trajectory interface}] that produces the specific chirping profile of Eq.~\eqref{eq: Linear chirp} in a spacetime diagram. The obtained interface profile, despite the complexity of the mathematical solution in Eq.~\eqref{eq: Trajectory interface}, is not very complex because the chirp specification in Eq.~\eqref{eq: Linear chirp} is fairly simple (linear chirp). However, the method is applicable to chirping profiles of arbitrary complexity. 

% \section{Conclusion}
% \pati{Summarize} Conclude that the scattering coefficient formulas work.
% \\
% \pati{Future Work} Arbitrary accelerated slab, superluminal accelerated interface, \dots.

\vspace{-1cm}
%%% References
{\small
% \setstretch{0.9}

}


\begin{thebibliography}{10}
% \setlength{\itemsep}{-0.01ex}
\setlength{\itemsep}{-1ex}

% \bibitem{<name>}
% e.g. 
% \bibitem{paper}
% C.R. Simovski and S.A. Tretyakov, ``Local constitutive parameters of metamaterials from an effective-medium perspective," {\itshape Physical Review B,} vol. 75, p. 195111, 2007.

\bibitem{Caloz2019a}
C. Caloz and  Z.-L. Deck-L\'eger, ``Spacetime metamaterials--Part I: General concepts," {\itshape IEEE Trans. Antennas Propag.,} vol. 68, no. 3, pp. 1569--1582, 2019.

\bibitem{Caloz2019b}
C. Caloz and  Z.-L. Deck-L\'eger, ``Spacetime metamaterials--Part II: Theory and applications," {\itshape IEEE Trans. Antennas Propag.,} vol. 68, no. 3, pp. 1583--1598, 2019.

\bibitem{Caloz2022}
C. Caloz, Z.-L. Deck-L\'eger, A. Bahrami, O. C. Vicente, and Z. Li, ``Generalized space-time engineered modulation (GSTEM) metamaterials: A global and extended perspective.'' {\itshape IEEE Antennas Propag Mag.,} 2022.

\bibitem{Doppler1842}
C. Doppler, ``\"{U}ber das farbige {L}icht der {D}oppelsterne und einiger anderer {G}estirne des {H}immels,'' {\itshape K{\"o}nigl. B{\"o}hm Gedsellsch. d. {W}is.,} vol 2, pp. 465--482, 1842.

\bibitem{Fresnel1818}
A. Fresnel, ``Lettre d'{A}ugustin {F}resnel \`{a} {F}ran\c{c}ois {A}rago sur l'influence du mouvement terrestre dans quelques ph\'{e}nom\`{e}nes d'optiques,'' {\itshape Ann. Chim. Phys.,} vol 9, pp. 57--66, 1818.

\bibitem{Huidobro2019}
P. A. Huidobro, E. Galiffi, S. Guenneau, R. V. Craster, and J. Pendry, ``Fresnel drag in space-time-modulated metamaterials,'' {\itshape Proc. Natl. Acad. Sci.,} vol. 116, no. 50, pp. 24 943--24 948, 2019.

\bibitem{Deck-Leger2022}
Z.-L. Deck-L\'eger, N. Chamanara, M. Skorobogatiy, M. G. Silveirinha, and C. Caloz, ``Uniform-velocity spacetime crystals,'' {\itshape Adv. Photonics,} vol. 1, no. 5, p. 056002, 2019.

\bibitem{bahrami2022}
A. Bahrami, Z.-L. Deck-L\'eger, and C. Caloz, ``Electrodynamics of accelerated-modulation space-time metamaterials," {\itshape Physical Review Applied,} vol. 19, p.054044, 2023.

\bibitem{Bahrami2023}
A. Bahrami, Z.-L. Deck-L\'eger Z. Li, and C. Caloz, ``Generalized FDTD scheme for moving electromagnetic structures with arbitrary space-time configurations,'' {\itshape arXiv preprint arXiv:2306.10035,} 2023.

\bibitem{bladel2012}
J. Van Bladel, {\itshape Relativity and Engineering,} Springer Science \& Business
Media, 2012, vol. 15.

\bibitem{Saleh2019}
B. E. A. Saleh and M. C. Teich, {\itshape Fundamentals of Photonics,} 3th edition, New York: Wiley, 2019.



\end{thebibliography}
\end{document}